\documentclass[12pt]{article}
\pdfoutput=1

\usepackage{amsmath,amssymb,amscd}
\usepackage{listings}
\usepackage{caption}
\usepackage{dsfont}
\usepackage{slashed}
\usepackage{color}
\usepackage{ulem}

\usepackage[pdftex]{graphicx}
\usepackage{epstopdf}
\usepackage{subfigure}
\usepackage{epsfig}
\usepackage{listings}
\usepackage{caption}
\usepackage{cite}

\usepackage{multirow}

\newcommand{\beq}{\begin{equation}}
\newcommand{\eeq}{\end{equation}}
\newcommand{\nbea}{\begin{align*}}
\newcommand{\neea}{\end{align*}}
\newcommand{\nbeq}{\begin{equation*}}
\newcommand{\neeq}{\end{equation*}}

 \usepackage{multirow}
\usepackage{array}
\newcolumntype{M}[1]{>{\centering\arraybackslash}m{#1}}
\newcolumntype{N}{@{}m{0pt}@{}}

\begin{document}


\pagestyle{empty}

\baselineskip=21pt
\rightline{{\fontsize{0.40cm}{5.5cm}\selectfont{KCL-PH-TH/2018-37, CERN-TH/2018-166}}}
\vskip 0.3in

\begin{center}

{\large {\bf Limits on Neutrino Lorentz Violation from Multimessenger Observations of TXS 0506+056}}

\vskip 0.3in

{\bf John Ellis}\textsuperscript{a,b,c},~
{\bf Nikolaos E. Mavromatos}\textsuperscript{a,d},~\\
{\bf Alexander S. Sakharov}\textsuperscript{e,f},~
{\bf Edward K. Sarkisyan-Grinbaum}\textsuperscript{f,g},~\\

\vskip 0.2in

{\small {\it

\textsuperscript{a}Theoretical Particle Physics and Cosmology Group, Physics Department \\
King's College London, Strand, London WC2R 2LS, United Kingdom\\
\vspace{0.25cm}
\textsuperscript{b}Theoretical Physics Department, CERN, CH-1211 Gen\`eve 23, Switzerland\\
\vspace{0.25cm}
\textsuperscript{c}NICPB, R{\"a}vala pst. 10, 10143 Tallinn, Estonia  \\
\vspace{0.25cm}
\textsuperscript{d}Currently also at: Department of Theoretical Physics and IFIC, University of Valencia - CSIC, Valencia, E-46100, Spain  \\
\vspace{0.25cm}
\textsuperscript{e}Physics Department, Manhattan College\\
{\mbox 4513 Manhattan College Parkway, Riverdale, NY 10471, United States of America}\\
\vspace{0.25cm}
\textsuperscript{f}Experimental Physics Department, CERN, CH-1211 Gen\`eve 23, Switzerland \\
\vspace{0.25cm}
\textsuperscript{g}Department of Physics, The University of Texas at Arlington\\
{\mbox 502 Yates Street, Box 19059, Arlington, TX 76019, United States of America} \\}}

\vskip 0.2in

{\bf Abstract}

\end{center}

\baselineskip=18pt \noindent


{\small
The observation by the IceCube Collaboration of a high-energy ($E \gtrsim 200$~TeV) neutrino from the direction of the
blazar TXS 0506+056 and the coincident observations of enhanced $\gamma$-ray emissions from
the same object by MAGIC and other experiments can be used to set stringent constraints on Lorentz
violation in the propagation of neutrinos that is linear in the neutrino energy: $\Delta v = - E/M_1$,
where $\Delta v$ is the deviation from the velocity of light,
and $M_1$ is an unknown high energy scale to be constrained by experiment.
Allowing for a difference in neutrino and photon propagation times of $\sim 10$~days, we find that
$M_1 \gtrsim 3 \times 10^{16}$~GeV.
This improves on previous limits on linear Lorentz violation in neutrino propagation by many orders of magnitude,
and the same is true for quadratic Lorentz violation.}


\vskip 0.5in

\leftline{November 2018}

\newpage
\pagestyle{plain}

It is desirable to probe fundamental physical principles as sensitively as possible,
and Lorentz invariance is no exception. Specifically, one may ask how accurately
we know that different species of massless particles travel at the speed of light, and
how accurately we know that massive particles travel at the same speed in the
high-energy limit. Over the past two decades, since the publication of~\cite{AEMNS},
considerable effort has been put by many experimental collaborations  into constraining different forms of Lorentz
violation, and specifically a linear coefficient $M_1$ in the velocity $v$ of
energetic photons: $\Delta v = - E/M_1$, using distant time-dependent astrophysical sources of energetic photons
such as pulsars, gamma-ray bursts (GRBs) and active galactic nuclei (AGNs). However,
analyses of possible Lorentz violation in photon propagation have been beset by difficulties in
disentangling intrinsic time delays in the sources from time delays accumulated during
propagation, and we consider that the strongest robust limit on $M_1$ for photons
is between $10^{17}$ and $10^{18}$~GeV~\cite{Rosti}. There have also been analyses of possible
Lorentz violation in neutrino propagation from Supernova 1987A and in a terrestrial neutrino
beam, but these are sensitive only to $M_1 \sim 2 \times 10^{11}$~GeV and potentially $\sim 4 \times 10^8$~GeV,
respectively~\cite{Harries}. More recently, data on the first observed black-hole binary merger~\cite{LIGO}
were used to to set the much weaker limit $M_1 \gtrsim 100$~keV for graviton
propagation~\cite{EGN}, and the near-coincidence of gravitational waves and $\gamma$-rays
from a neutron-star binary merger has been used to establish that their velocities are the same
to within $\sim 10^{-17}$~\cite{LIGOetal}.

Very recently, the IceCube Collaboration has reported the observation of an ultra-high-energy
neutrino from the direction of the blazar TXS 0506+056, and together with a number of other groups,
most notably the MAGIC Collaboration, have reported~\cite{IceCube} an enhanced level of activity in
$\gamma$-ray and photon emission from this source, which is located at a distance
$\sim 4 \times 10^9$~ly. As we discuss in this paper, the great distance of TXS 0506+056 and the high energy $\gtrsim 200$~TeV
of the observed high-energy neutrino, in conjunction with the $\gamma$-ray observations,
provides unique sensitivity to Lorentz violation in neutrino propagation, which almost rivals
that to linear Lorentz violation in photon propagation~\footnote{For a previous test of Lorentz violation assuming that IceCube neutrino IC~35~\cite{IC35}
was emitted by a flare of the AGN PKS B1424-418~\cite{Kadler}, see~\cite{WLW}. In that case, the chance coincidence probability was $\sim 5$\%,
so the identification could not be considered conclusive. We
note also that a correlation between a flaring $\gamma$-ray source and the IceCube-160731
neutrino event was reported in~\cite{AGILE}, but it was not possible to identify the potential counterpart and
make a quantitative analysis.}. The
sensitivity to linear Lorentz violation in neutrino propagation is to
$M_1 \gtrsim 3 \times 10^{16}$~GeV, approaching
the Planck energy scale that might be characteristic of the possible quantum-gravity
effects that were the original motivation for~\cite{AEMNS}.

We first review the observations of TXS 0506+056 reported by the IceCube Collaboration and the teams
studying its electromagnetic emissions~\cite{IceCube}. The primary observation by IceCube was that of a
single neutrino with energy $\sim 290$~TeV (90\% CL lower limit 183~TeV) on 22 September 2017,
dubbed IceCube-170922A, coming from a direction within $0.1^o$ of the catalogued $\gamma$-ray
source TXS 0506+056, whose redshift $z = 0.3365 \pm 0.0010$.
Several $\gamma$-ray experiments, notably MAGIC, VERITAS, HESS, {\it Fermi}-LAT,
{\it AGILE} and Swift made observations showing that TXS 0506+056 was in a flaring state over a period
within about 10 days of IceCube-170922A~\cite{IceCube}. In particular, MAGIC reported a 6.2-$\sigma$ excess
within this time frame. The IceCube Collaboration has also reported an excess of
neutrinos observed earlier from the direction of TXS 0506+056, confirming this as the source of
IceCube-170922A~\cite{IceCube2}, and analyses have supported the hypothesis that a single astrophysical
mechanism is responsible for emitting both the neutrino and the $\gamma$-rays~\cite{Follow-ups}.

The similarity in arrival times of IceCube-170922A and the electromagnetic emissions can be used
immediately to estimate the corresponding sensitivity to a difference $\Delta v_{\nu \gamma}$ in the propagation
speeds {\it in vacuo} of the neutrino and photons, assuming that both speeds are independent of energy.
We assume a distance of $4 \times 10^9$~ly and an illustrative time difference of 10~days~\footnote{The
redshift of TXS 0506+056 is not very large, and the estimates of $\Delta t$ and the energy of
the neutrino are not very accurate, so this estimate does not include the small effects associated with the expansion of the Universe during propagation.}, so that
$\Delta v_{\nu \gamma}/c \sim$ 10~days/$4 \times 10^9$~years $\sim 10^{-11}$~\footnote{Henceforth,
we use natural units in which the conventional velocity of light $c = 1$.}. This is six orders of magnitude worse
than the corresponding constraint on the difference in propagation speeds of gravitational waves and
photons derived from the near-simultaneous observations of the binary neutron-star merger:
$\Delta v_{GW \gamma} \lesssim 10^{-17}$~\cite{LIGOetal}. However, it is much better than
the corresponding sensitivity to an energy-independent $\Delta v_{\nu \gamma}$
from the observations of neutrinos emitted during the collapse of supernova 1987A:
$\Delta v_{\nu \gamma} \lesssim$ 4~hours/$1.5 \times 10^5$~years $\sim 3 \times 10^{-9}$~\footnote{We
note that a Fermi all-sky variability analysis reported significant brightening of TXS 0506+056 in the GeV band some five
months previous to the observation of IceCube-170922A~\cite{IceCube}. A conservative approach would be to allow for a time difference of
150 days between the photon and neutrino propagation times, which would relax our bound on $\Delta v_{\nu \gamma}$ by
a factor $\simeq 15$. However, it would still be over an order of magnitude stronger than the bound from supernova 1987A.}.

An energy-independent difference between the velocities of neutrinos (or gravitational waves)
and photons would require the extremely radical step of abandoning the framework of special relativity.
A less radical hypothesis would be that Lorentz invariance is an emergent symmetry in the low-energy
limit, but is subject to modification that increases with energy. This is indeed the suggestion that has been
made in a number of different theoretical frameworks, including the `space-time foam' expected in
models of quantum gravity~\cite{Wheeler}, phenomenological models suggested by
features of cosmic-ray physics~\cite{mestres} and other considerations~\cite{pheno},
the suggestion that Lorentz invariance may be broken
spontaneously~\cite{sme,pospelov}, models of loop quantum gravity~\cite{loop}, doubly-special relativity theories~\cite{dsr}
and quantum field theories of the Lifshitz type~\cite{lifsh}. In such frameworks,
Lorentz invariance is a good symmetry in the low-energy limit, but is violated increasingly at high energies.
As discussed in~\cite{EMN}, the interactions of particles with space-time foam are not represented by an effective field theory
(EFT) with higher-dimensional operators such as the standard model extension~\cite{sme}, since they correspond to
time-space uncertainty effects. Since other models of Lorentz violation may also not fall within an EFT framework, we
take here a phenomenological approach in which the energy dependence of Lorentz violation is kept free, and the
magnitude is allowed to be different for different particle species.

The first such possibility that we consider is that $\Delta v_{\nu \gamma}$ increases linearly with energy:
$\Delta v_{\nu \gamma} = - E/M_1$~\footnote{Constraints on $e^+ e^-$ pair production {\it in vacuo}
require $\Delta v < 0$~\cite{CG}, as expected in the model of~\cite{EMN}.}. The possibility of such a linear violation of Lorentz invariance was
raised in~\cite{AEMNS,EMN} on the basis of intuition about the properties of space-time foam
suggested by a heuristic string-inspired model of quantum-gravitational fluctuations in space-time.
In such a case, one's first guess could be that $M_1$ would be comparable to the Planck mass:
$M_1 \sim M_P \simeq 10^{19}$~GeV. However, the value of $M_1$ would depend in a string-inspired model on unknown
quantities such as the string coupling, the density of defects in space-time, and the strength of
particle interactions with such defects, which may not be universal between different particle
species~\cite{EMS}, so we maintain
phenomenological open minds about the possible magnitude of $M_1$. The model of space-time
foam proposed in~\cite{EMN} would suggest that the velocities of neutrinos would deviate from
the low-energy velocity of light less than photons, so that (in an obvious notation)
$M_{1, \nu} \gg M_{1, \gamma}$, because the photon would have stronger interactions with the space-time
defects. This is because, in such a stringy model of space-time foam,
only species that carry no non-trivial quantum numbers under the standard model group have unsuppressed interactions with the foam,
in which case the fact that neutrinos are fermions with non-trivial SU(2)$_L$ properties renders space-time foamy effects
invisible to them. However, initially we will be agnostic whether the photon velocity or the neutrino
velocity deviates more from the low-energy velocity of light. When they are comparable,
$M_1 = (M_{1, \gamma} \times M_{1, \nu})/(M_{1, \gamma} - M_{1, \nu})$, but when there is a hierarchy
between them, $M_1 \to$ the smaller of $M_{1, \gamma}$ and $M_{1, \nu}$.

We recall that a difference in velocity $\Delta v = - E/M_1$ induces a difference in arrival time
$\Delta t = \Delta v \times D = (E \times D)/M_1$, where $D$ is the propagation distance.
For our numerical purposes, we assume the value $E_\nu = 200$~TeV for the energy of the event
IceCube-170922A~\cite{IceCube}, and note that the energies of the $\gamma$-rays measured by MAGIC
and other experiments are negligible in comparison. A simple order-of-magnitude estimate then yields a sensitivity to
\begin{equation}
M_1
\gtrsim
 \frac{H_0^{-1}}{\Delta t}E
\int\limits_0^{z_{\rm src}}
\frac{(1+z)}{\sqrt{\Omega_{\Lambda}+\Omega_M(1+z)^3}}dz \approx 3 \times 10^{16}~{\rm GeV} \,  ,
\label{M1Limit}
\end{equation}
which is over 6 orders of magnitude stronger than the limit obtained previously~\cite{Harries} from an
analysis of the neutrino signal from supernova 1987A~\footnote{In calculating (\ref{M1Limit}) we used the standard cosmological
$\Lambda$CDM model with dark energy and dark matter contributions $\Omega_{\Lambda} = 0.7$ and  $\Omega_M = 0.3$,
respectively, and  Hubble expansion rate $H_0=68$~km/s/Mpc. See~\cite{Rosti} for detailed derivation of (\ref{M1Limit}).}.
The sensitivity (\ref{M1Limit}) is, nevertheless, an order of magnitude weaker than the robust limit on photon Lorentz violation~\cite{Rosti},
so refers directly to the neutrino.

It is instructive also to compare the sensitivity (\ref{M1Limit}) to
the possible improvement in the supernova limit, should another core-collapse supernova be observed in our galaxy.
Multi-dimensional simulations of such events suggest that their neutrino emissions might exhibit time
variations in the millisecond range, in which case measurements might attain a sensitivity to $M_1 \sim 2 \times 10^{13}$~GeV~\cite{Janka},
still 3 orders of magnitude less than the IceCube-170922A/MAGIC sensitivity (\ref{M1Limit}). This sensitivity is
also far beyond that we can envisage using a terrestrial neutrino beam. It was estimated using the timing capabilities of
the OPERA detector and assuming that timing information could be available for neutrino events upstream in rock that
a sensitivity to $M_1 \sim 4 \times 10^8$~GeV could be attained~\cite{Harries}~\footnote{In fact, we are unaware
of neutrino experiments that have sought to test Lorentz invariance in the way proposed here. For alternative
searches for Lorentz violation using neutrinos, see~\cite{MINOS,SK,T2K}, see also~\cite{Stecker}. We are grateful to Francesca Di Lodovico,
Brian Rebel and Jenny Thomas for discussions on this subject.}. Thus the IceCube-170922A/MAGIC sensitivity
seems to outclass the capabilities of terrestrial experiments as well as possible future supernova observations.

One can also consider a possible quadratic violation of Lorentz invariance: $\Delta v = - E^2/M_2^2$,
which would be an option in some of the alternative models of Lorentz violation mentioned
above~\cite{mestres,pheno,sme,pospelov,loop,lifsh}. In this case, the IceCube-170922A/MAGIC sensitivity
would be to
\begin{equation}
M_2
 \gtrsim
 \left[\frac{3}{2}\frac{H_0^{-1}}{\Delta
t}E^2\!\!
\int
\limits_0^{z_{\rm src}}\!\!
\frac{(1+z)^2}{\sqrt{\Omega_{\Lambda}+\Omega_M(1+z)^3}}dz\right]^{1/2}
\!\!\!\!
 \approx 10^{11}~{\rm GeV}
\, ,
\label{M2Limit}
\end{equation}
which is over 5 orders of magnitude stronger than the corresponding limit from supernova 1987A~\cite{Harries}.
In the case of quadratic Lorentz violation, the supernova 1987A limit was estimated to be to $M_2 \sim 4 \times 10^4$~GeV, the possible sensitivity of a
future galactic supernova event was estimated to be to $M_2 \sim 10^6$,
and the potential sensitivity of a terrestrial experiment was estimated to be to $M_2 \sim 7 \times 10^5$~GeV~\footnote{For
the most sensitive terrestrial measurement of neutrino propagation speed, see~\cite{MINOSToF}.}.
Again, the large distance of TXS 0506+056 and the high energy of the IceCube-170922A event enable it to
outclass the competition.

For completeness, before closing we comment briefly on previous discussions of neutrino Lorentz violation
in the context of EFT and the Standard Model Extension (SME)~\cite{sme}.
This has been mentioned~\cite{Stecker2} as an explanation of a possible drop in PeV neutrinos
suggested by IceCube data~\cite{drop}, which might correspond to a SME dimension-6 term with
coefficient $\ge - 5.2 \times 10^{-35}$$~{\rm GeV^{-2}}$ (see also~\cite{1708} for a review). However, we regard the existence of this drop and its
interpretation as questionable. An overview of this and other aspects of SME applications to neutrinos
is given in~\cite{1609}, though without a quantitative discussion.

We conclude that the advent of multimessenger neutrino/photon astronomy~\cite{IceCube,IceCube2}
has not only launched a new era in the study of the origins of high-energy cosmic rays, but also made
possible a breakthrough in the exploration of Lorentz symmetry using neutrinos. We may anticipate that
more coincidences between high-energy neutrino events and electromagnetic emissions will be
observed, enabling the rough estimates made here to be refined and improved. Such coincidences
would contribute to fundamental physics as well as resolving important issues in astrophysics.

\section*{Acknowledgements}

The research of J.E. and N.E.M. was supported partly by the STFC Grant ST/L000258/1.
N.E.M. also acknowledges the hospitality of IFIC Valencia through a
Scientific Associateship ({\it Doctor Vingulado}).  The work of A.S.S.
was supported partly by the US National Science Foundation under Grants
PHY-1505463
and PHY-1402964.



\end{document}